\documentclass[conference]{IEEETran}
\usepackage{sansmath}
\usepackage{url}
\usepackage{cite}
\usepackage{amsmath,amssymb,amsfonts}
\usepackage{algorithm}
\usepackage[noend]{algpseudocode}
\usepackage{graphicx}
\usepackage{textcomp}
\usepackage{tabularx}
\usepackage{booktabs}
\usepackage{xcolor}
\usepackage{xspace}
\usepackage[prefixflatinterpret,prefixflatbracketmodalinterpret,prefixflatsetinterpret,sidenotecalculus,longseqcontext,silentconst]{logic}
\usepackage[prefixflatinterpret,prefixflatbracketmodalinterpret,differentialdL,simplenames,silentconst]{dL}
\usepackage{hyperref}
\newcommand{\seq}{\,;\,}

\usepackage{enumitem}

\usepackage{amsthm}
\theoremstyle{definition}

\usepackage{xcolor}

\usepackage{float}
\floatstyle{ruled}
\newfloat{model}{tbp}{model}
\floatname{model}{Model}
\newcounter{modelline}
\makeatletter
\newcommand{\mline}[1]{{\refstepcounter{modelline}\ltx@label{#1}}~\text{\scriptsize{\themodelline}}\quad}
\makeatother

\usepackage{prettyref}
\newcommand{\rref}[2][]{\prettyref{#2}}
\newrefformat{sec}{Section\,\ref{#1}}
\newrefformat{subsec}{Section\,\ref{#1}}
\newrefformat{def}{Def.\,\ref{#1}}
\newrefformat{thm}{Theorem\,\ref{#1}}
\newrefformat{prop}{Proposition\,\ref{#1}}
\newrefformat{lem}{Lemma\,\ref{#1}}
\newrefformat{cor}{Corollary\,\ref{#1}}
\newrefformat{ex}{Example\,\ref{#1}}
\newrefformat{tab}{Table\,\ref{#1}}
\newrefformat{fig}{Fig.\,\ref{#1}}
\newrefformat{app}{Appendix\,\ref{#1}}
\newrefformat{alg}{Algorithm\,\ref{#1}}
\newrefformat{model}{Model\,\ref{#1}}
\newrefformat{line}{Line\,\ref{#1}}
\newrefformat{eq}{Eq.\,\eqref{#1}}
\newrefformat{case}{Case\,\eqref{#1}}
\newrefformat{appendix}{Appendix\,\ref{#1}}

\newcommand{\kwd}[1]{{\normalfont \textsf{#1}}\xspace} 
\newcommand{\avg}{\kwd{avg}}
\newcommand{\vavg}{\kwd{vavg}}
\newcommand{\sol}{\frac{2v_0 (v_1 + v_2)}{2v_0 + v_1 + v_2}}

\newcommand{\artifactdoi}{\url{https://doi.org/10.1184/R1/28934195}}

\begin{document}

\title{Can Large Language Models Autoformalize Kinematics?}

\author{
\IEEEauthorblockN{Aditi Kabra\IEEEauthorrefmark{1}, Jonathan Laurent\IEEEauthorrefmark{2}, Sagar Bharadwaj\IEEEauthorrefmark{1},\\
Ruben Martins\IEEEauthorrefmark{1}, Stefan Mitsch\IEEEauthorrefmark{3}, Andr\'e Platzer\IEEEauthorrefmark{2}}
\IEEEauthorblockA{\IEEEauthorrefmark{1}Carnegie Mellon University, Pittsburgh, USA}
\IEEEauthorblockA{\IEEEauthorrefmark{2}Karlsruhe Institute of Technology, Karlsruhe, Germany}
\IEEEauthorblockA{\IEEEauthorrefmark{3}DePaul University, Chicago, USA}
\IEEEauthorblockA{\{akabra, skalasib, rubenm\}@cs.cmu.edu, jonathan.laurent@kit.edu, smitsch@depaul.edu, platzer@kit.edu}
}

\maketitle

\begin{abstract}
Autonomous cyber-physical systems like robots and self-driving cars could greatly benefit from using formal methods to reason reliably about their control decisions. However, before a problem can be solved it needs to be stated. This requires writing a formal physics model of the cyber-physical system, which is a complex task that traditionally requires human expertise and becomes a bottleneck.

This paper experimentally studies whether Large Language Models (LLMs) can automate the formalization process. A 20 problem benchmark suite is designed drawing from undergraduate level physics kinematics problems. In each problem, the LLM is provided with a natural language description of the objects' motion and must produce a model in differential game logic (\dGL). The model is (1) syntax checked and iteratively refined based on parser feedback, and (2) semantically evaluated by checking whether symbolically executing the \dGL formula recovers the solution to the original physics problem.
A success rate of 70\% (best over 5 samples) is achieved. We analyze failing cases, identifying directions for future improvement. This provides a first quantitative baseline for LLM-based autoformalization from natural language to a hybrid games logic with continuous dynamics.
\end{abstract}


\section{Introduction}
\label{sec:introduction}

Formal methods seem especially useful in the safety-critical but mathematically complex domain of cyber-physical systems (CPSs), i.e., systems like robots and planes where discrete software interacts with the continuous dynamics of the real world.
However, a bottleneck preventing broader industry adoption is the challenge of writing a formalization for these systems \cite{10.1007/978-3-319-48869-1_2}.
Unlike program verification, where a mathematically meaningful object to verify already exists in the form of a program, for CPSs, we must first create a formal model of the physical system and environment.
This process is notoriously difficult, time-consuming, and error-prone.

Can large language models (LLMs) help autoformalize physical systems?
They have shown autoformalization abilities for mathematics \cite{10.5555/3600270.3602614,NEURIPS2024_6034a661} and CPS contracts \cite{chen2023nl2tl,10161125}.
This paper presents a first, experiment-focused exploration of the ability of LLMs to autoformalize the underlying physics problems, faithfully preserving their exact continuous dynamics and discrete transitions in differential game logic (\dGL) \cite{DBLP:journals/tocl/Platzer15}.
We propose 20 benchmarks derived from hard undergraduate kinematics problems \cite{irodov1988,mitocwpset,ling_sanny_moebs_2016}.
We find that, supported by few-shot prompting and parser feedback, OpenAI o3 has a top-5 accuracy of 70\% (success rate on selecting the best out of 5 attempts per benchmark).
The main cause of failure is checker limitations for the most complex problems.

Our benchmarks and pipeline are available as an online artifact at \artifactdoi.
Evaluating autoformalization of a physical system is challenging because every problem has many valid formal representations, and conversely, models can be incorrect for many subtle reasons.
To automatically assess the semantic proximity of generated formal models to the original physics problem, we provide a checker that:
\begin{enumerate}
\item performs \emph{symbolic execution} over the generated formal model, examining whether the expected symbolic solution to the original physics problem is recovered.
\item ensures the model is not in stasis by requiring that a minimum number of variables are mutated, indicating they are subject to physical effects.
\end{enumerate}
Recovering the expected solution to the original natural language problem is a strong indication that the formalization models the problem correctly, though the checker may sometimes reject answers that are in principle correct but missed some implicit assumptions that the original solution made.

Autoformalization of physical models can have tremendous impact on reliable autonomy.
Unlike in mathematics, validating a hypothesized CPS formalization is possible with comparisons between real measurements and predicted measurements during test runs \cite{DBLP:journals/fmsd/MitschP16}.
Easy formalization would allow every modular component to be modeled formally, with all assumptions explicit.
Verification could lead to safer systems and synthesis could make control system design easier while ensuring no edge cases are missed relative to the formal model.

\section{Overview}

This section explains by example what the process of autoformalizing a physics problem involves.
Consider the following physics question \cite[problem 1.2]{irodov1988}, which will serve as a running example:

\begin{quote}
    ``A point traversed half the distance with a velocity \(v_0\). The remaining part of the distance was covered with velocity \(v_1\) for half the time, and with velocity \(v_2\) for the other half of the time. Find the mean velocity of the point averaged over the whole time of motion."\\
    \textit{Solution: }``\(v_{\avg} = \sol\)"
\end{quote}

The physical motion described here is hybrid, with continuous dynamics (motion at velocity \(v_0\), \(v_1\), and finally \(v_2\)) as well as discrete transitions (changing velocity under specified conditions).
Additionally, there are non-trivial conditions describing \emph{when} the discrete transitions happen.
We seek to symbolically model this motion formally.

A canonical way to express control problems is via hybrid games \cite{10.5555/646734.701483,DBLP:journals/IEEE/TomlinLS00}.
Hybrid games are commonly formalized in two ways: via hybrid automata \cite{10.5555/646734.701483} and via logic \cite{DBLP:journals/tocl/Platzer15}.
In this paper, we target the logic formalization as it is closer to code, which LLMs have been more exposed to and trained for.
\emph{Differential game logic} (\dGL) is a logic expressing two-player hybrid games with a relatively complete axiomatization \cite{DBLP:journals/tocl/Platzer15}.
A detailed and gradual introduction can be found in the literature \cite{Platzer18}.
\rref{model:running-example} shows an example of a \dGL hybrid game which formalizes the running example.

\begin{model}[h]
  \setcounter{modelline}{0}
  \caption{Running Example: Finding Mean Velocity}
  \label{model:running-example}
  \vspace{-\baselineskip}
  \begin{align*}
    \text{\kwd{setup}} &\,\big| \mline{line:setup} \big\langle \, x := 0 \seq d_{\vavg} := 0 \seq \\
    \text{\kwd{phase 0}} &\,\big| \mline{line:phase0} \phantom{\big\langle \,} \left\{ x' = v_0, d_{\vavg}' = v_{\avg} \right\} \seq \phantom{\big[} d_h := x \seq \\
    \text{\kwd{phase 1}} &\,\big| \mline{line:phase1} \phantom{\big\langle \,} t := 0 \seq \left\{ x' = v_1, d_{\vavg}' = v_{\avg}, t' = 1 \right\} \seq \\
    \text{\kwd{phase 2}} &\,\big| \mline{line:phase2} \phantom{\big\langle \,} t_h := t \seq \left\{ x' = v_2, d_{\vavg}' = v_{\avg}, t' = 1 \right\} \seq \\
    \text{\kwd{transitions}} &\,\big| \mline{line:check} \phantom{\big\langle \,} \ptest{t = 2t_h} \seq \ptest{x = 2d_h} \\
    \text{\kwd{win when}} &\,\big| \mline{line:wincondition} \big\rangle \, d_{\vavg} = x
  \end{align*}
\end{model}

In this game, a player (canonically called Angel) runs physical dynamics per the problem specification, moving the point's position \(x\) with velocity \(v_0\) in phase 0 on \rref{line:phase0}, with velocity \(v_1\) in phase 1 on \rref{line:phase1}, and with velocity \(v_2\) in phase 2 on \rref{line:phase2}).
Angel is forced to transition between phases under the exact conditions specified by the problem because of \emph{tests} (assertions) on \rref{line:check}.
These assertions use \emph{auxiliary variables} \(t_h\) to keep track of the time elapsed in phase 1, \(t\) to keep track of time elapsed in phase 1 and phase 2 combined, and \(d_h\) to keep track of half the distance covered.
On \rref{line:check}, test \(\ptest{x = 2d_h}\) ensures that the first phase transition occurred when half the distance was covered.
The test \(\ptest{t = 2t_h}\) ensures that the second transition happened at half the total run time of phases 1 and 2.
If Angel fails either test condition, she immediately loses the game.

Notice that there is a variable \(v_{\avg}\) representing average velocity, and \rref{line:wincondition} says that Angel will win the game only if \(v_{\avg}\) was correct, as checked using auxiliary variable \(d_{\vavg}\), which tracks the displacement covered by a particle traveling at \(v_{\avg}\) through all phases.
The value that \(v_{\avg}\) must hold for Angel to win matches the solution to the original problem, \(v_{\avg} = \sol\).
In this example, there happens to be no adversarial dynamics (Angel is the only player making decisions).
However, in problems with adversarial nondeterminism such as unpredictable environmental factors or controller latency, a second, symmetric, adversarial player (canonically called Demon) models the situation by resolving nondeterminism with the goal to make Angel lose.
Adversarial dynamics are also useful for optimization problems, e.g., ``choose the launch angle that minimizes drift", which can be modeled as Angel and Demon competing to find this angle.

Having formalized the problem as a \dGL hybrid game, we next describe how symbolic execution enables a check for whether the model aligns with the natural language question.
For the fragment of \dGL where most of the chosen benchmark problems lie (loop-free, polynomial solutions), symbolic execution is decidable and existing work \cite{DBLP:conf/tacas/KabraLMP24} describes how to perform it.
Backwards symbolic execution over a game lets us compute the conditions under which Angel (or dually, Demon) can win the game.
For example, symbolic execution of the subgame \(\langle \rref{line:check} \rangle d_{\vavg}=x\) at the end of \rref{model:running-example} evaluates to \(x=d_{\vavg} \land t=2t_h \land x=2d_h\) which is precisely the weakest precondition starting from which Angel passes the tests of \rref{line:check} and ends the game in a state where she wins.
Consider the precondition \(\phi\) of \rref{line:setup} computed in this way.
Angel should be able to win the game exactly when \(v_{\avg}\) is set to the correct solution, i.e., \(v_{\avg}=\sol \leftrightarrow \phi\).
Such a check is implemented using Z3 and Mathematica.

Observe that even a minor modeling mistake can result in this check failing.
In fact, symbolically executing \rref{model:running-example} reveals that it is actually equivalent to \(\top\).
It suffers from a subtle exploit where Angel can always run each phase for 0 time, in which case \(x_\avg = x = 0\) regardless of what \(v_\avg\) is set to.
The problem is that the model lets Angel choose the distance that the particle will travel, as her choice at \rref{line:phase0} determines the value of half-distance \(d_h\), which in turn determines the distance that phases 1 and 2 must cover.
The fix is simple: the assignment \(d_h := x\) on \rref{line:phase0} should be turned to a test \(\ptest{d_h = x}\).
The automated check was able to catch this difference and is a strong signal of correctness, especially in combination with manual review.
The automated checker has a second test to ensure the system is not in stasis, ruling out empty games that exploit the symbolic execution test by setting the win condition directly to the desired solution.
For this example, the second test checks via static analysis that at least two variables are written to, since \emph{any} reasonable model will at least modify the variables representing the particle's position and time.
The checker is provided with the minimum number of variables that must be written to and the expected solution for every benchmark problem.

Finally, we see the benefit of formal modeling when the model with \rref{line:phase0} corrected is still not equivalent to the expected solution.
The problem is that the textbook solution is correct \emph{only under some unstated assumptions}: distance \(d\) is non-zero, velocity \(v_0\) is positive, and \(v_1+v_2\) is also positive.
Finally, \rref{model:fixed}, makes these assumptions.
Under the assumptions of \rref{model:fixed}, \rref{line:assum}, the modal formula from \rref{model:fixed}, \rref{line:setup2} to \rref{line:wincondition2} is equivalent to \(\sol\).
In unverified or informal physics, subtle modeling glitches are widespread.
As a human, it is easy to neglect a measure zero case like \(d_h = 0\) during an intermediate calculation, but formalization forces this case to be explicit.

\begin{model}[h]
  \setcounter{modelline}{0}
  \caption{Repaired Model: Finding Mean Velocity}
  \label{model:fixed}
  \vspace{-\baselineskip}
  \begin{align*}
    \text{\kwd{assum2}} &\,\big| \mline{line:assum} (d_h>0 \land v_0>0 \land v_1+v_2>0) \limply \\
    \text{\kwd{setup}} &\,\big| \mline{line:setup2} \big\langle \, x := 0 \seq d_{\vavg} := 0 \seq \\
    \text{\kwd{phase 0}} &\,\big| \mline{line:phase02} \phantom{\big\langle \,} \left\{ x' = v_0, d_{\vavg}' = v_{\avg} \right\} \seq \phantom{\big[} \ptest{d_h = x} \seq \\
    \text{\kwd{phase 1}} &\,\big| \mline{line:phase12} \phantom{\big\langle \,} t := 0 \seq \left\{ x' = v_1, d_{\vavg}' = v_{\avg}, t' = 1 \right\} \seq \\
    \text{\kwd{phase 2}} &\,\big| \mline{line:phase22} \phantom{\big\langle \,} t_h := t \seq \left\{ x' = v_2, d_{\vavg}' = v_{\avg}, t' = 1 \right\} \seq \\
    \text{\kwd{transitions}} &\,\big| \mline{line:check2} \phantom{\big\langle \,} \ptest{t = 2t_h} \seq \ptest{x = 2d_h} \\
    \text{\kwd{win when}} &\,\big| \mline{line:wincondition2} \phantom{\big[} \big\rangle \, d_{\vavg} = x
  \end{align*}
\end{model}

This example begins to demonstrate why formalizing a hybrid physical problem is \emph{not} a simple translation task.
Introducing the right auxiliary variables, absent in the specification, is often necessary for a clean, tractable formalization.
All the pieces of the model must fit together precisely, e.g., even making a weak inequality strict can transform a correct model into one admitting no runs \cite{TAN2021247}.
Correctness requires careful attention to the subtleties of the problem specification,
e.g., benchmark problem 10 (available in the online artifact~\cite{artifact}) has one variable representing a \emph{displacement} and another, representing a \emph{distance}.
The former is a signed number while the latter must be modeled as non-negative.
Models must prune mathematically degenerate cases using background knowledge of the physical world, e.g., by eliminating imaginary roots of polynomials computing real quantities.
Additionally, there are many ways to model the same problem, and the right choice of coordinate system or frame of motion can be crucial to writing a model for which symbolic reasoning is tractable.

Given the complexity of the task, it is unclear whether LLMs can succeed at all, making a careful evaluation necessary.
We create a benchmark suite drawing from textbooks and problem sets \cite{irodov1988,mitocwpset,ling_sanny_moebs_2016}, choosing tricky problems with varying structures to systematically identify the strengths and weaknesses of LLMs on autoformalization.

\section{Methodology}

\begin{figure}[h]
    \centering
    \includegraphics[width=\linewidth]{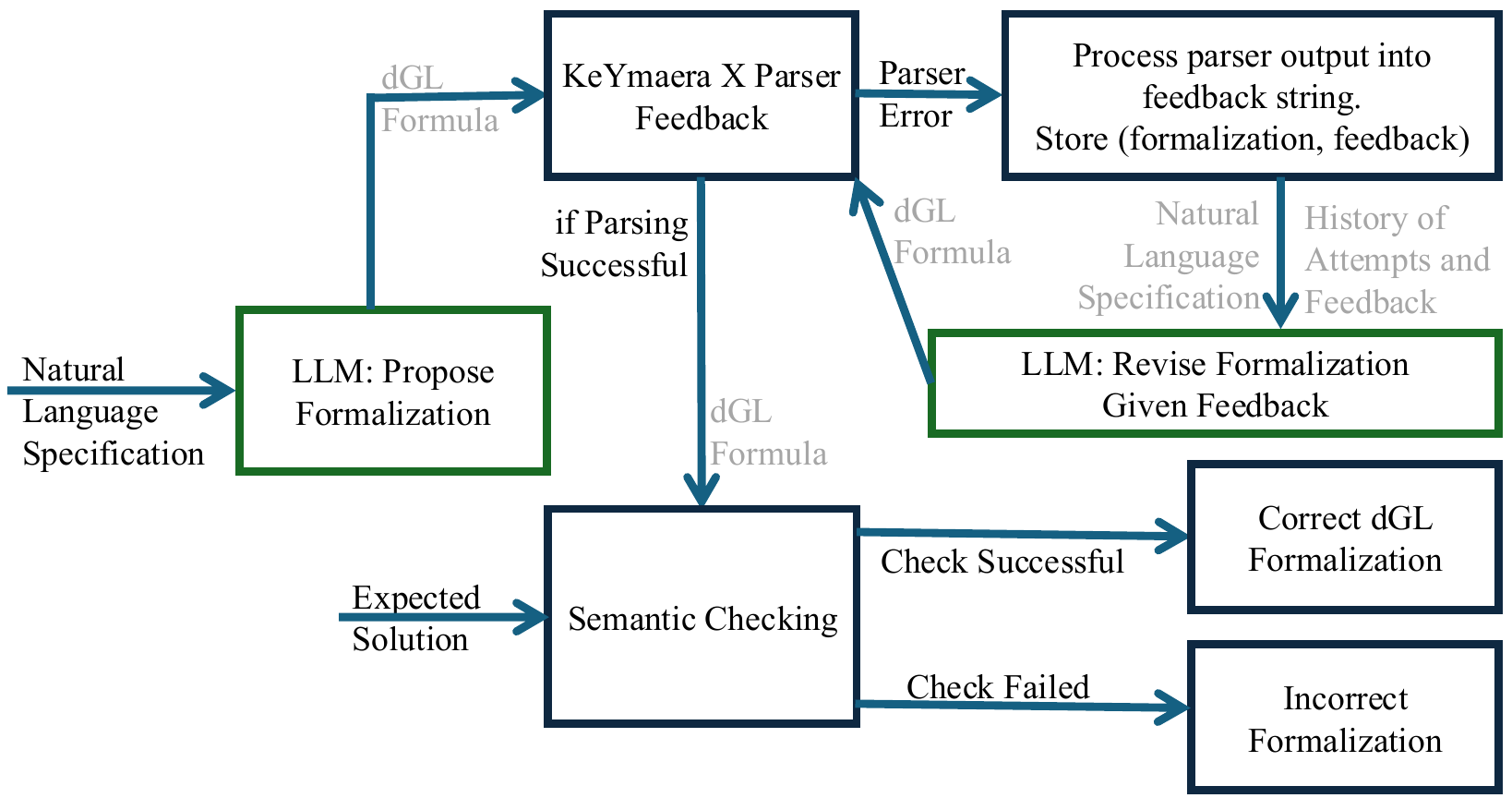}
    \caption{Autoformalization pipeline}
    \label{fig:pipeline}
\end{figure}

The autoformalization system, shown by \rref{fig:pipeline}, takes as input the natural language specification of the physical problem to model, and produces either a correct \dGL formalization of a failure (no formalization found that passed the checker).
It uses two types of LLM queries: one is \emph{Propose Formalization}, and the second one is \emph{Revise Formalization}.
The artifact~\cite{artifact} shows all the prompt templates used.

The autoformalization system first queries the LLM to propose the formalization using the \emph{Propose Formalization} query, providing up to four solved examples as part of the prompt.
It provides an LLM with the natural language description of the problem and asks it to produce a \dGL formula, including in the prompt common \dGL syntax mistakes and their remedies.
An example of a problem description the prompt would use based on the running example is
\begin{quote}
    ``An object A starts from rest and travels a distance d. For the first half of the distance, it travels with velocity v0. For the second half of the distance, it spends half the time traveling with velocity v1 and half the time traveling with velocity v2. Another object B tracks A. Its motion has the same duration, starting, and ending points as A, but it travels with uniform velocity vavg. Let vavg remain a free variable."
\end{quote}
The prompt also includes upto four solved examples.
The returned formal model is passed to the KeYmaera X \dGL parser for a syntax check.

If the syntax check fails, then the parser output is processed into short feedback strings.
An example of a parser feedback string is
\begin{quote}
        ``The input formula contains an unsupported Unicode character (possibly \texttt{\{bad\_char\}}). Use only ASCII characters."
\end{quote}
Where \texttt{bad\_char} is a parameter extracted from parser output.
The incorrect formalization along with the feedback is then appended to a list of previous failed formalization attempts.

The LLM is prompted to repair the formalization given past attempts and feedback using the \emph{Revise Formalization} query.
The prompt first provides up to two examples of repair.
Then it provides the natural language question, and finally past incorrect proposals along with the feedback indicating why these proposals were wrong.
The LLM replies with a new, repaired \dGL game, which is again passed to the parser for checking.
This syntactic repair loop is allowed to run at most three times, after which autoformalization fails.
A full implementation, along with examples, is available in the online artifact~\cite{artifact}.

If the syntax check succeeds, then the \dGL formula is passed onto the semantic evaluation system.
This system takes as input the expected symbolic solution to the problem along with the proposed \dGL formula.
For our running example, the expected solution consists of the formula \(v_{\avg} = \sol\), to test against for equivalence, and minimum expected number of variables that are written to, 2.
If the checker finds that the \dGL formula is equivalent to the expected solution under the initial assumptions of the formula and has enough variable writes, then it declares that the formalization is acceptable.
The pipeline is implemented using Delphyne \cite{delphyne} and \KeYmaeraX \cite{DBLP:conf/cade/FultonMQVP15}, which in turn uses Z3~\cite{DBLP:journals/corr/DeMouraM08} and Mathematica.

\section{Related Work}

Given a natural language description, formal physics models are generally written manually. Prior work has studied the use of structured natural language to perform formalization by rule-based transformations \cite{Carvalho2016,8256284}, but this limits flexibility in input specification. Recent progress in NLP (LLMs) raises the prospect of automating formalization directly from unstructured natural language.

The use of LLMs for autoformalizing mathematics has received significant research attention \cite{10.5555/3600270.3602614,NEURIPS2024_6034a661}, including for purely continuous systems with partial differential equation \cite{huang-etal-2025-llms,du2024large}.
Recent work also uses LLMs to synthesize contracts for cyber-physical systems \cite{chen2023nl2tl,10161125,Neider2025}.
However, we focus on LLM based autoformalization for \emph{physics models} of hybrid systems, with discrete transitions and continuous dynamics faithfully represented, which to the best of our knowledge has not yet been studied.
Compared to general mathematics and LTL specifications, we expect this domain to be more data-scarce, with few examples of differential dynamic logic system formalizations and even fewer examples of differential game logic games available on the Internet.
Additionally, we expect CPS model formalization to have a more operational flavor, modeling physical realities required to make a system work rather than abstract concepts that can be simplified for theoretical convenience (mathematics) or focused on only output behavior (LTL contracts).
This difference makes it worth investigating the hybrid systems modeling domain.

\section{Evaluation}

We propose 20 benchmarks derived from challenging problems in undergraduate physics textbooks and problem sets \cite{irodov1988,mitocwpset,ling_sanny_moebs_2016}.
The online artifact~\cite{artifact} lists them.
Phrasing is changed to focus on modeling the physical situation described in the problem (rather than on solving for a variable) and the numbers are changed to induce rational answers to avoid failures of the automated check over numerical imprecision.
Each benchmark is accompanied by an expected symbolic solution that the model should be equivalent to and the minimum expected number of variables that must be written to.
Four additional problems, separate from the benchmarks, serve as examples for few-shot prompting.

We run the LLM autoformalization system using OpenAI GPT 4o, GPT 4.1 (best available OpenAI non-reasoning model) and OpenAI o3 (best available OpenAI reasoning model) on the benchmark set, getting successively improving results.
Outcomes are classified into 4 buckets:
\begin{enumerate}
    \item \emph{Success}: Correct, checker-certified formalization.
    \item \emph{Failed}: autoformalization failed either the syntax or semantic check.
    \item \emph{Timeout}: The solver timed out after 3 minutes of symbolic execution.
    Symbolic execution uses real quantifier elimination, which is doubly exponential \cite{DBLP:journals/jsc/DavenportH88}, and can result in timeouts. 
    \item \emph{Tool Failure}: The solver encountered an expression that it could not automatically symbolically execute (e.g., ODE with a square roots on the right hand side).
\end{enumerate}

\begin{figure}
    \centering
    \includegraphics[width=\linewidth]{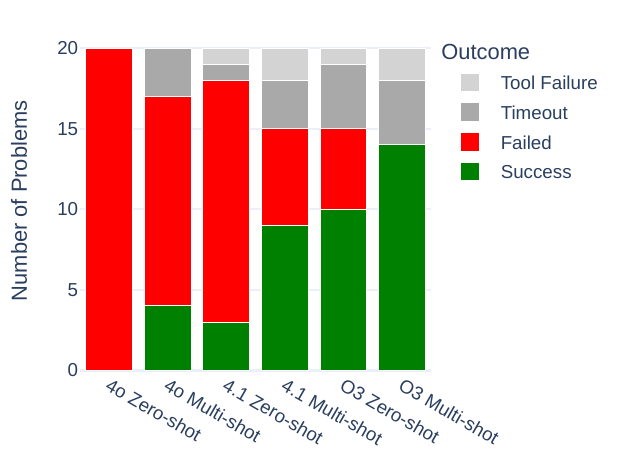}
    \caption{Outcomes of autoformalization by GPT 4o, GPT-4.1 and o3, with zero-shot and multi-shot prompting.}
    \label{fig:results}
\end{figure}

\rref{fig:results} shows the outcome. The autoformalization system is sampled five times independently with temperature 1, and the best outcome is reported (where tool failure and solver timeout are considered better than failure).
For each sample, the LLM is given three chances to repair syntax if the \KeYmaeraX parser reported an error. We evaluate two different prompts, one using few-shot prompting considering four examples (\emph{multi-shot}) and another one without examples (\emph{zero-shot}).

GPT 4o without few-shot prompting fails on every benchmark.
Few-shot prompting significantly improves performance for all models. With few-shot prompting o3 has a success rate of 70\%, which is higher than GPT 4.1's 45\% and GPT 4o's 20\%. Our results support that the improvements and the introduction of reasoning LLM models significantly increase their success rate on mathematical tasks, enabling strong performance in the autoformalization of hybrid games.

\begin{figure}
    \centering
    \includegraphics[width=1.0\linewidth]{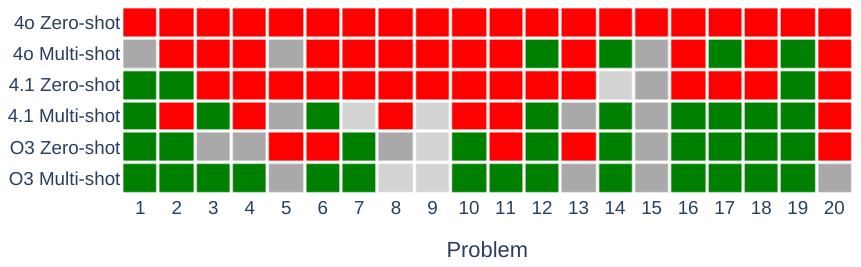}
    \caption{Problem outcome by model.}
    \label{fig:heatmap}
\end{figure}

\section{Discussion}
\rref{fig:heatmap} shows the outcome of each problem by model. We can observe that models found the same problems hard.
There are six problems that no model solves.
For these, the best performing model produces a formalization that could potentially be correct but is too complex for the checker to verify.
The problems can be classified into three groups that explain where their complexity arises from.
\begin{enumerate}
    \item Optimization Problems: Problem 5 and 20.
    \item Non-polynomial continuous dynamics: Problem 8 and 9.
    \item Complicated dynamics: Problem 13 and 15.
\end{enumerate}
The first category, for example, includes problem 5, which is about identifying the launch angle of a boat that minimizes drift, modeled by adversarial Angel trying to pick a better angle than Demon, who must set the optimal angle to win.
The interplay of the two players leads to combinatorial branching over possible sequence of events, while continuous dynamics occur under constraints preventing control blocking.
These constraints make symbolic reasoning about the differential equations particularly computationally expensive.

An example of the second category is problem 8, which is about drones arranged in an equilateral triangle that cyclically pursue each other with constant radial velocity, collapsing inwards in increasingly rapid revolutions, till they crash.
Their dynamics, often modeled with a non-polynomial square root expression on the right side of an ODE, cannot be handled automatically by the tool.
Problem 13 is an example from the third category. It has three phases of dynamics, including one where acceleration has a continuous dependence on time.
Such a differential equation is more expensive to symbolically evaluate, making a timeout the likely outcome.

Sometimes a problem fundamentally requires capabilities that a solver lacks, and the solver must be extended.
Other times a difficult but crucial part of formalizing problems is choosing the right abstractions and representations that let proofs succeed.
For the six failing problems, the solution likely lies in a combination of these methods.
The solver must be extended to better automatically handle, for instance, non-polynomial dynamics by using continuous invariant generation \cite{DBLP:journals/fmsd/SogokonMTCP22}.
Complementing this, the LLM should be guided to use tricks like polar coordinates and auxiliary variables to rewrite dynamics in friendlier forms, and to also provide checking hints such as continuous invariants.

Unsolved benchmarks provide a roadmap for improvement in both these directions.
Additionally, the rapid improvement in success rate displayed by better LLM models suggests that autoformalization of hybrid games will continue to improve with advancements in LLMs.

\section{Conclusion and Future Work}

This work experimentally studies the ability of LLMs to formally model physical behavior given a natural language specification.
The results (70\% success rate) are encouraging, and suggest that CPS model autoformalization using LLMs could be an impactful direction for future research.

Beyond the broader adoption of formal methods in offline controller design, physics autoformalization enables a new class of applications, where autonomous systems autoformalize unfamiliar situations they face in the open world and derive formally justified control decisions to respond to them.
For such applications, where the \emph{physical environment} must be formalized at runtime, the input to autoformalization would be vision and sensor data rather than a natural language description.
LLM autoformalization from such multimodal input data provides an interesting future research direction.

\section*{Acknowledgment}
This work was partially supported by the National Science Foundation (NSF) under Award Number CCF2427581, DARPA under Agreement FA8750-24-9-1000, and by an Alexander von Humboldt Professorship.

\section*{Authors' Note}
This is a pre-print. The definitive version of record appears in the proceedings of FMCAD 2025 (\url{https://fmcad.org}), as:

\noindent
Aditi Kabra, Jonathan Laurent, Sagar Bharadwaj, Ruben Martins, Stefan Mitsch, André Platzer. ``Can Large Language Models Autoformalize Kinematics?”, FMCAD 2025.

\bibliographystyle{plain}
\bibliography{refs}

\end{document}